\documentclass[twocolumn,superscriptaddress,showpacs,nofootinbib,notitlepage]{aastex631}
\usepackage{graphicx,subfigure,amsmath,amsfonts,amssymb,multirow,mathrsfs,ulem,threeparttable}
\usepackage{hyperref}
\usepackage{epstopdf}
\begin{document}
\title{Probing Cosmology with 92 Localized Fast Radio Bursts and DESI BAO}

\correspondingauthor{Yi-Zhong Fan}
\email{yzfan@pmo.ac.cn}

\author[0000-0003-1215-6443]{Yi-Ying Wang}
\affiliation{Key Laboratory of Dark Matter and Space Astronomy, Purple Mountain Observatory, Chinese Academy of Sciences, Nanjing 210033, People's Republic of China}

\author[0000-0002-0822-0337]{Shi-Jie Gao}
\affiliation{School of Astronomy and Space Science, Nanjing University, Nanjing, 210023, People's Republic of China}
\affiliation{Key Laboratory of Modern Astronomy and Astrophysics, Nanjing University, Ministry of Education, Nanjing, 210023, People's Republic of China}

\author[0000-0002-8966-6911]{Yi-Zhong Fan}
\affiliation{Key Laboratory of Dark Matter and Space Astronomy, Purple Mountain Observatory, Chinese Academy of Sciences, Nanjing 210033, People's Republic of China}
\affiliation{School of Astronomy and Space Science, University of Science and Technology of China, Hefei, Anhui 230026, People's Republic of China}

\newcommand{\ud}{\mathrm{d}}
\begin{abstract}
Recent baryon acoustic oscillation (BAO) measurements from the Dark Energy Spectroscopic Instrument (DESI) collaboration, combined with the cosmic microwave background (CMB) and type Ia supernovae (SNe Ia) observations, suggest a preference for dynamical dark energy (DDE) with $w_0>-1$ and $w_a<0$. Given the cosmological origin of fast radio bursts (FRBs), the combination of their dispersion measures and host galaxy redshifts makes localized FRBs a valuable tool for probing cosmology. Using an updated sample of 92 localized FRBs, along with DESI BAO, PlantheonPlus and CMB data, we constrain the dark energy (DE) equation of state (EoS) under the Chevallier-Polarski-Linder (CPL) parameterization. We find that even without incorporating CMB data, DDE remains preferred with $w_0 = -0.855 ^{+0.084}_{-0.084}$ and $w_a = -1.174^{+0.462}_{-0.491}$ at a confidence level of $\sim2.5 \sigma$. A joint analysis constrains these to be $w_0 = -0.784^{+0.064}_{-0.064}$ and $w_a = -0.872^{+0.269}_{-0.278}$, showing a discrepancy with $\Lambda$CDM at a $\sim3.1\sigma$ level. Furthermore, using localized FRBs alone, we estimate the Hubble constant $H_0$ to be $69.04^{+2.30}_{-2.07}$ and $75.61^{+2.23}_{-2.07} \, \rm km \, s^{-1} \, Mpc^{-1}$, assuming the Galactic electron density models to be NE2001 (Cordes \& Lazio) and YMW16 (Yao et al.), respectively. Thus, accurate accounting of the Galactic dispersion measure is crucial for resolving the Hubble tension with FRBs. Future BAO measurements, next-generation CMB experiments, and more localized FRBs will further constrain the DE EoS and the cosmological parameters. 
\end{abstract}

\section{Introduction}\label{sec:1}
Standing as the cornerstone in modern cosmology, the cosmological constant $\rm \Lambda$ and the cold dark matter (CDM) model ($\rm \Lambda$CDM) explains the majority of the cosmological observations successfully. However, certain signals in cosmological and astrophysical data manifest significant tensions with the Planck18 parameters values \citep{2020A&A...641A...6P} or standard cosmology model, e.g., the Hubble tension, the $S_8$ tension, CMB anisotropy anomalies and small-scale curiosities \cite[for reviews, see][]{2021CQGra..38o3001D, 2022JHEAp..34...49A, 2022NewAR..9501659P, 2023Univ....9...94H}. These tensions are often interpreted as potential indicators of new physics beyond the standard $\rm \Lambda$CDM model. It is crucial to measure the cosmological parameters through other independent astrophysical observations.

As tracers of large-scale structure, galaxies, quasars and Lyman-$\alpha$ forests reveal the feature of baryon acoustic oscillations (BAO), providing a key probe of the cosmic expansion history. A recent BAO measurement from the Dark Energy Spectroscopic Instrument (DESI) shows a preference for dynamical dark energy (DDE) at a confidence level of  $\sim 2.5 \sigma$, $3.5 \sigma$, and $3.9  \sigma$ \citep{2024arXiv240403002D}, when combined with CMB plus Pantheon+, Union3, and DES-SN5YR supernova datasets, respectively. Compared to the cosmological constant $\rm \Lambda$, DDE not only offers a potential partial resolution to the Hubble tension \citep{2017NatAs...1..627Z, 2020PhRvD.101l3516A} but also provides a natural explanation for the surprisingly abundant massive galaxies at extremely high redshift \citep{2023JCAP...10..072A, 2024JCAP...07..072M, 2024EPJP..139..711W} observed with James Webb Space Telescope \citep{2023Natur.616..266L}. However, there is ongoing debate over whether a robust preference for DDE \citep{2024arXiv240917074G, 2024arXiv240716689G}.

Fast radio bursts (FRBs) can serve as a promising cosmic probe for exploring the equation of state (EoS) of DE, as firstly proposed by \citet{2014PhRvD..89j7303Z} (a specific FRB/Gamma-ray burst association scenario was also discussed by \citealt{2014ApJ...788..189G}). With typical radiation frequencies around GHz, these transient signals are significantly dispersed by the ionized medium along their propagation path, which is quantified by the dispersion measure (DM).
Since FRBs are from cosmological distances, the DM caused by the inhomogeneous intergalactic medium (IGM) dominates, which in principle could provide valuable information of Hubble parameter $H(z)$ and DE \citep[e.g.,][]{2014PhRvD..89j7303Z,2021SCPMA..6449501X, 2022A&ARv..30....2P, 2023RvMP...95c5005Z}.
In particular, FRB and BAO data can offer complementary constraints of the matter density parameter $\Omega_m$ and the DE EoS parameter $w$, making their combination a powerful tool to investigate the nature of DE \citep{2014PhRvD..89j7303Z}. Thanks to the advancements in radio instrumentation and a relatively high burst rate of FRBs (a rate of a few thousand per day, \citealt{2024Ap&SS.369...59L}), 843 FRBs (including 67 repeating FRBs) have been detected\footnote{\url{https://blinkverse.zero2x.org/}} \citep{Xu+2023}. With these growing datasets, statistical analyses and cosmological applications are becoming feasible \citep{2024arXiv240913247W}.

Localized FRBs have been employed to measure the Hubble constant $H_0$: the information of the luminosity distance can be extracted from the dispersion measure $\rm DM_{IGM}$ contributed by the IGM; the confirmed host galaxies provide accurate redshift measurements.
Using a sample of 69 localized FRBs, \citet{2024arXiv241003994G} obtained the tightest constraint on $H_0 = 70.41^{+2.28}_{-2.43} \rm \, km \, s^{-1} \, Mpc^{-1}$ with $\Omega_m$ and the baryon density parameter $ \Omega_b h^2$ following the combined results from CMB+BAO \citep{2020A&A...641A...6P}.
Similarly, \cite{2024arXiv241001974K} derived $H_0 = 70.42^{+3.73}_{-3.63} \rm \, km \, s^{-1} \, Mpc^{-1}$ from 64 localized FRBs, where the DM contribution from the host galaxy ($\langle \rm DM_{host} \rangle$) was treated as a model parameter. With the increasing number of localized FRBs, an independent measurement of $H_0$ in the local Universe are expected to resolve the Hubble tension.

In this work, we collect 92 localized FRBs and will constrain the DE EoS using a combination of DESI BAO, PantheonPlus \citep{2022ApJ...938..113S} and CMB observations. The DE EoS is parameterized using the Chevallier-Polarski-Linder (CPL) model, expressed as $w(a) = w_0 + w_a (1-a)$ \citep{2001IJMPD..10..213C, 2003PhRvL..90i1301L}, where the model reduces to the standard $\rm \Lambda$CDM scenario when $w_a=0$ and $w_0=-1$. We will explore both the flat $w$CDM and $w_0 w_a$CDM models and test their preferences. Additionally, leveraging the increased number of FRB events, we will independently measure $H_0$ from these localized FRBs, representing almost the tightest constraints compared with previous works. This paper is organized as follows. In \autoref{sec:2}, we build the likelihood functions of FRBs, DESI BAO and PantheonPlus. In \autoref{sec:3}, we perform the Bayesian analytic results, including the DDE model and Hubble constant. In \autoref{sec:4}, we present conclusions and summary.

\section{Methods}\label{sec:2}
Considering a flat universe ($\Omega_{\rm K} = 0$), the Hubble parameter $H(z)$ in $w_0 w_a$CDM model is
\begin{widetext}
\begin{equation}
H(z) = H_0 \sqrt{\Omega_m (1+z)^3 + {\rm \Omega_R}(1+z)^4 + {\rm \Omega_{DE}}(1+z)^{3(1+ w_0 + w_a)} \mathrm e^{-3 w_a z/(1+z)} }, 
\end{equation}\label{eq:1}
\end{widetext}
where $w_0$ represents the present-day DE EoS, $w_a$ quantifies its dynamical evolution, $\Omega_m$, $\rm \Omega_R$ and $\rm \Omega_{DE}$ are the matter, radiation and dark energy density parameters, respectively.  We use $D_{\rm M}/r_{\rm d}$, $D_{\rm H}/r_{\rm d}$ and $D_{\rm V}/r_{\rm d}$ from the measurements of DESI BAO \citep{2024arXiv240403002D}. These terms correspond to different distance measures, including the comoving distance $D_{\rm M}(z)$, the equivalent distance variable $D_{\rm H}(z)$ and the angle-average distance $D_{\rm V}(z)$, which can be written as
\begin{equation}\label{eq:2}
\begin{aligned}
D_{\rm M}(z) & = \int^{z}_{0} \frac{c \, {\rm d}z'}{H(z')}, \\
D_{\rm H}(z) & = \frac{c}{H(z)}\\
{\rm and~} D_{\rm V}(z) & = \big(zD_{\rm M}(z)^2 D_{\rm H}(z) \big)^{1/3}.
\end{aligned}
\end{equation}
The sound horizon at the drag epoch is defined as $r_{\rm d} = \int^{\infty}_{z_{\rm d}} {c_{\rm s}(z)}/{H(z)} {\rm d} z$ ($z_{\rm d} \simeq 1060$), where $c_{\rm s}$ is the speed of sound in the primordial plasma. For FRB measurements, $H(z)$ directly affects the IGM contribution to the DM \citep{2014PhRvD..89j7303Z}, given by
\begin{equation}\label{eq:3}
\langle {{\rm DM_{IGM}}(z)} \rangle = \frac{3 c  \Omega_b H^2_0 f_{\rm IGM}}{8\pi G m_p} \int^{z}_0 \frac{\chi(z')(1+z') {\rm d}z'}{H(z')},
\end{equation}
where $\Omega_b$ is the baryon density parameter, $m_p$ is the proton mass, $f_{\rm IGM} \simeq 0.84$ is the fraction of baryons in the IGM and $\chi(z')=7/8$ accounts for the ionization fractions of intergalactic fully ionized hydrogen and helium \citep[e.g.,][]{2012ApJ...759...23S, 2020MNRAS.496L..28L}. Since the observed DM ($\rm DM_{\rm obs}$) can be split into multiple terms, the IGM component can be derived by \citep{2014ApJ...783L..35D}
\begin{equation}
{\rm DM_{IGM}} = {\rm DM_{obs}} - {\rm DM_{MW}} - {\rm DM_{halo}} - \frac{\rm DM_{host}}{1+z},
\end{equation}
where $\rm DM_{MW}$, $\rm DM_{halo}$ and $\rm DM_{host}$ are the contributions from the Milky Way, the Galactic halo and the host galaxies, respectively. $\rm DM_{MW}$ is mainly produced by the contribution of warm ionized medium and can be calculated by the Milky Way electron density models, e.g., NE2001 \citep{2002astro.ph..7156C} and YMW16 \citep{2017ApJ...835...29Y}. In this work, we adopt NE2001 model by default for $\rm DM_{MW}$ and discuss the impact of different models in \autoref{sec:3}. 

\citet{2019MNRAS.485..648P} suggested that $\rm DM_{halo} = 50-80 \, pc \, cm^{-3}$ for integration up to the virial radius $r_{200}$, within which the average density equals $200\rho_c$, where $\rho_c$ is the critical density. As a result, studies in the literature usually treated the Galactic halo contribution ($\rm DM_{halo}$) as a constant ($\sim 50 \rm \, pc \, cm^{-3}$) \citep{2022MNRAS.516.4862J, 2024MNRAS.527.7861G, 2024arXiv240703532F} or following a uniform or Gaussian distribution \citep{2022MNRAS.515L...1W, 2024arXiv241001974K}. In this work, we adopt a more detailed physical model for $\rm DM_{halo}$ as constructed by \citet{2019MNRAS.485..648P}. In this model, the dark matter halo is described by a modified NFW profile \citep{2017ApJ...846L..24M} with $M_{\rm halo}=1.5\times10^{12} \,M_{\odot}$ \citep{2020MNRAS.494.5178F}. The mass of the diffuse baryons in the halo is $M_{b, {\rm halo}} = f_{b, {\rm halo}}M_{\rm halo}(\Omega_b / \Omega_m)$, where $f_{b, {\rm halo}} =0.75$ specifies the mass fraction. Therefore, for a specific cosmological framework, the baryon density $\rho_{b}$ can be 
derived and the electron density is $n_{\rm e} = \mu_{\rm e}{\rho_{ b}}/({m_{p} \mu_{\rm H}})$. Here, $\mu_{\rm H} = 1.3$ is the reduced mass (accounting for helium) and $\mu_{\rm e}=1.167$ accounts for fully ionized helium and hydrogen. Since the calculations of $r_{200}$, the mass of the baryons, and the $\rm DM_{halo}$ profile depend on different cosmological parameters, the values of $\rm DM_{halo}$ are also influenced by the cosmological framework. For our FRB samples, we integrate $n_{\rm e}$ from $0$ to $3r_{200}$ to estimate the contribution of $\rm DM_{halo}$ \citep[e.g.,][]{2019MNRAS.485..648P}. In this case, the halo contribution can be evaluated for any Galactic coordinates and is more reliable when estimating cosmological parameters.

Unlike the well-established understanding of the Galactic distribution of free electrons, $\rm DM_{host}$ connects with the physical origin of FRBs and the characteristic of their host galaxy, which still remains less well constrained. \citet{2020Natur.581..391M} adopted a log-normal distribution to model this component: 
\begin{equation}
p_{\rm host}({\rm DM_{host}|\mu, \sigma_{\rm host}}) = \frac{1}{\sqrt{2\pi}{\rm DM}\sigma_{\rm host}} {\rm exp} \left[ - \frac{({\rm ln \, DM} - \mu)^2}{2\sigma_{\rm host}^2} \right],
\end{equation}
where a correction ${\rm DM_{host}} \to {\rm DM_{host}}/{(1+z)}$ has been applied. The median value is $\mathrm e^{\mu}$ and the  standard variance is $\mathrm e^{\mu+\sigma^2_{\rm host}/2}(\mathrm e^{\sigma^2_{\rm host}-1})^{1/2}$. 
In statistical studies, \citet{2020ApJ...900..170Z} estimated $\rm DM_{host}$ for three types of FRBs from the IllustrisTNG simulation, including (1) the non-repeating FRBs, (2) the repeating FRBs localized to dwarf galaxies with stellar masses $\sim 1-50 \, \times 10^7 \, M_{\odot}$ and star formation rates $\sim 0.1-0.7 \, M_{\odot} \, \rm yr^{-1}$, and (3) the repeating FRBs that localized to spiral galaxies with stellar masses $\sim 0.1-10 \, \times 10^7 \, M_{\odot}$ and star formation rates $\sim 0.01-10 \, M_{\odot} \, \rm yr^{-1}$. Base on these criteria, we classify the localized FRBs accordingly and assume the value of $\rm DM_{host}$ following the distribution listed in Tab.3 of \citet{2020ApJ...900..170Z}.

\begin{figure}[ht!]
	\centering
	\includegraphics[width=\linewidth]{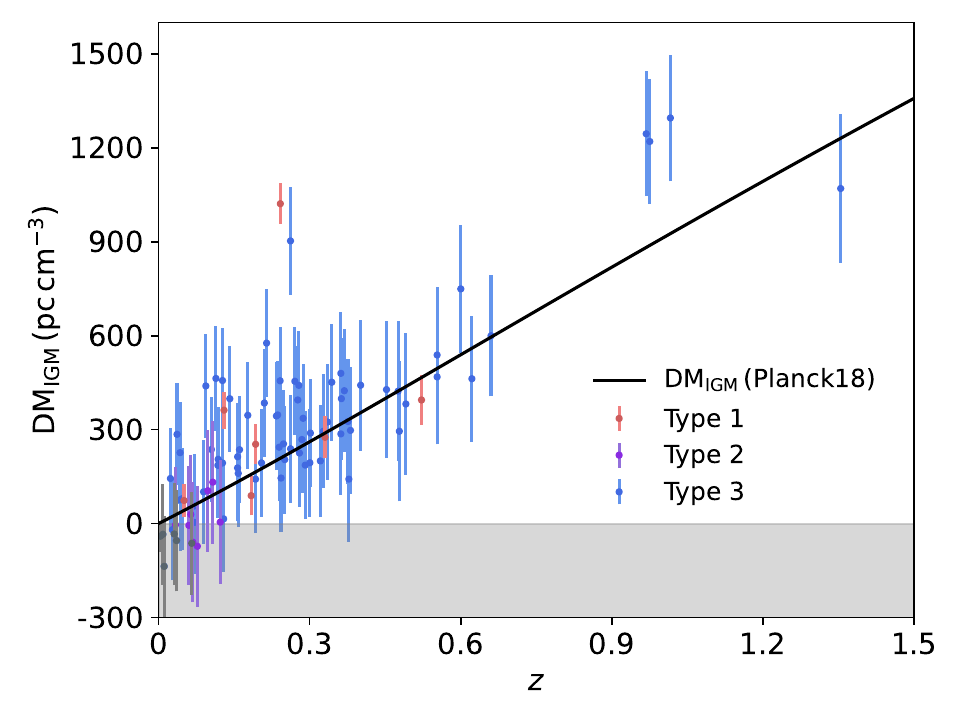}
	\caption{\small  The $\rm DM_{IGM}$-redshift relation for localized FRBs. The black solid line denotes the expected relation (\autoref{eq:3}) between $\rm DM_{IGM}$ and redshift based on Planck 2018 results \citep{2020A&A...641A...6P}. The colored data points represent the estimated $\rm DM_{IGM}$ and the redshift from their host galaxies. The criteria of classification for localized FRBs follows \citet{2020ApJ...900..170Z}. The mean values of $\rm DM_{IGM}$ are derived by subtracting the contributions of $\rm DM_{MW}$ (using NE2001 model), $\rm DM_{halo}$ and $\rm DM_{host}$. The corresponding errors are from the uncertainties of $\rm DM_{host}$ and fixed values of $30 \, \rm pc \, cm^{-3}$ for both $\rm DM_{MW}$ and $\rm DM_{halo}$. The grey data points  do not satisfy the additional criteria $\rm DM_{obs} - DM_{MW} > 80~{\rm pc~cm^{-3}}$  and are not used in analysis.}
	\label{fig:DMIGM-z}
\end{figure}

Since the components involved in the DESI BAO and FRBs measurements have been explicitly implemented, the likelihood functions can be established in the Bayesian statistical framework. Regarding \autoref{eq:2} as the model ($f(x)$), the likelihood functions for DESI BAO can be written as
\begin{equation}
\mathcal{L}_{\rm BAO}=\prod^{N}_{i} {\rm exp} \left[ -\frac{1}{2} \left(\frac{f(x_i)-y_i}{\sigma_i} \right)^2 \right],
\end{equation}
where $(x_i, y_i)$ and $\sigma_i$ are the measurements of DESI BAO and their uncertainties, respectively. For $\rm DM_{IGM}$, a quasi-Gaussian function with a long tail has been extensively used to build the likelihood function \citep{2020Natur.581..391M}, 
\begin{equation}
p_{\rm IGM}=A\Delta^{-\beta}{\rm exp} \left[ -\frac{(\Delta^{-\alpha} - C_0)^2}{2\alpha^2\sigma_{\rm DM}^2}\right], \quad \Delta>0,
\end{equation}
where $\Delta = {{\rm DM}_{\rm IGM}}/\langle {\rm DM}_{\rm IGM} \rangle$, $\alpha=\beta=3$ depict the inner density profile of gas in halos, and $\sigma_{\rm DM}$ is the effective standard deviation. Using the state-of-the-art IllustrisTNG cosmological simulation, \citet{2021ApJ...906...49Z} obtained the best-fit values of $A$, $C_0$ and $\sigma_{\rm DM}$ at different redshifts in the range of 0.1 -- 9. Here, we use the fitting results from Tab.1 of \citet{2021ApJ...906...49Z} to calculate $p_{\rm IGM}$. For all FRBs, the total likelihood function can be written as
\begin{widetext}
\begin{equation}
\mathcal{L}_{\rm FRB}= \prod^{N}_{i} \int^{{\rm DM_{obs, \,i} - DM_{MW, \, i} -DM_{halo, \, i}}}_{0} p_{{\rm host} , \, i}({\rm DM_{host}})p_{{\rm IGM}, \, i} {\rm d DM_{host}},
\end{equation}
\end{widetext}
where $N$ represents the total number of localized FRBs.
During Bayesian analysis, the distributions of $p_{\rm host}$ remain normalized at each integral interval before marginalization. An additional selection criteria, $\rm DM_{obs} - DM_{MW} -DM_{halo} > 0$, is applied to ensure the physical consistency of $\mathcal{L}_{\rm FRB}$. Considering the general contribution of $\rm DM_{halo}$ ($\sim 50 -80 \, \rm pc \, cm^{-3} $), we reselect FRBs by $\rm DM_{obs} - DM_{MW} > 80~{\rm pc~cm^{-3}}$.  This reselection can ensure a large prior range of parameter inference and minimize the bias of $\rm DM_{IGM}$ estimation as much as possible. Otherwise, $H_0$ will be constrained in a priori tight range with higher value, because these excluded FRBs always predict a smaller $\rm DM_{IGM}$. As shown in \autoref{fig:DMIGM-z}, this selection criteria can also exclude several low-redshift sources to avoid the bias from the peculiar velocities of galaxies.

Furthermore, we combine the likelihood functions from the datasets of PantheonPlus and Planck CMB (TT, TE, EE+lowE+lensing). For an observed SNe Ia with a confirmed redshift, the model distance modulus $\mu_{\rm model}$ is defined as
\begin{equation}
\mu_{\rm model}(z_i) = 5  {\rm log}_{10} \left[ \frac{(1+z)}{10 \, {\rm pc}}\int^{z_i}_{0}\frac{c{\rm d}z'}{H(z')} \right].
\end{equation}
Therefore, the likelihood can be calculated by $\chi^2$, which can be written as
\begin{equation}
\mathcal{L}_{\rm SNe} = {\rm exp}\bigg[-\frac{1}{2}{\rm \Delta}{\boldsymbol{D}}^{T} C^{-1}_{\rm stat+syst} {\rm \Delta}{\boldsymbol{D}} \bigg],
\end{equation}
where $\boldsymbol{D}$ is the vector of SNe Ia distance modulus residuals with ${\rm \Delta}D_i = \mu_i - \mu_{\rm model}(z_i)$, and $C_{\rm stat+syst}$ is the statistical and systematic covariance matrices \citep{2022ApJ...938..110B}. For the likelihood function $\mathcal{L}_{\rm CMB}$ of Planck CMB, we use the high-$\ell$ {\tt Plik\_lite} likelihood for the TT, TE and EE spectra, the low-$\ell$ {\tt commander} likelihood for the TT spectrum, the low-$\ell$ {\tt SimAll} likelihood for the EE spectra and the lensing likelihood from \citet{2020A&A...641A...5P, 2020A&A...641A...8P}. 

To balance efficiency and accuracy, we adopt {\tt Nessai} \citep{2024ascl.soft05002W} as the sampler and set 2000 live points during Bayesian analysis. As declared in \citet{2024arXiv240403002D}, the BAO distance ladder results can determine $\Omega_m$ and $H_0r_{\rm d}$, but the sound horizon is sensitive to the baryon density $\omega_b =  \Omega_b h^2$. Therefore, prior knowledge of $\omega_b$ is required to break the degeneracy between $H_0$ and $r_{\rm d}$. As shown in \autoref{eq:3}, an apparent degeneracy revealed between $\Omega_b$ and $H_0$. Based on a new measurement of the ${\rm D}(p, \, \gamma) {}^{3}{\rm He}$ cross section by the LUNA Collaboration,  \citet{2021JCAP...04..020P} reported a new measurement of $\Omega_b h^2=0.02233 \pm 0.00036$,  which was independent on the Planck results. In this case, we assume a Gaussian distribution as the prior of $\Omega_b h^2$. For the cases that include CMB likelihoods, $\Omega_b h^2$ is assumed to follow a wider uniform distribution in the range of $[0.02, 0.025]$ as listed in \autoref{Tab:priors}.

\begin{deluxetable}{cc}[ht!]
\centering
\tablecaption{The prior distributions of cosmological parameters \label{Tab:priors}}
\tablehead{\colhead{Parameter} &\colhead{Prior of parameter inference}}
\startdata
$H_0$ &Uniform(50, 90)\\
$\Omega_m$ &Uniform(0.1, 0.9)\\
$w_0$ &Uniform(-3, 0)\\
$w_a$ &Uniform(-3, 2)\\
$M$ &Uniform(-20, -18)\\
$\Omega_b h^2$ &Gaussian(0.02233, 0.00036) or Uniform(0.02, 0.025)\\
\enddata
\end{deluxetable}

\section{Results}\label{sec:3}
The localized FRBs used in this study are listed in \autoref{Tab:FRBs} (A machine-readable format is available in
the online journal and on \href{https://github.com/wangyy19/Localized-FRB-dataset}{GitHub}). FRB220208A, FRB220330D, FRB221027A, and FRB230216A are excluded, because the possibilities of their host associations are $\rm P_{host} < 90\%$ \citep{2024arXiv240916964S}.   \autoref{fig:DMIGM-z} shows the redshift and estimated $\rm DM_{IGM}$ values for our FRB samples. To constrain the flat $w_0 w_a$CDM and $w$CDM models, the parameter priors in \autoref{eq:1} and the fiducial magnitude $M$ for SNe Ia are listed in \autoref{Tab:priors}. The requirement $w_0 + w_a < 0$ is also imposed to enforce a period of high-redshift matter domination.

\begin{figure}[ht!]
	\centering
	\includegraphics[width=\linewidth]{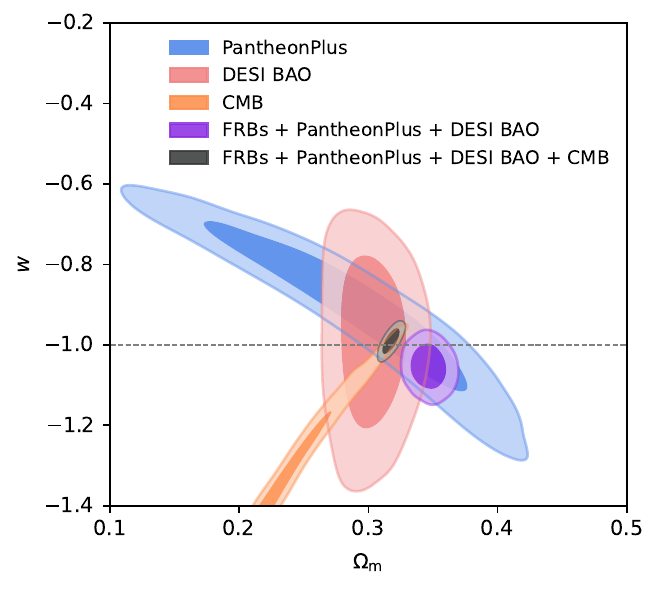}
	\caption{The posterior distributions of $\Omega_m$ and $w_0$ in the flat $w$CDM model. The dashed grey line represents the $\rm \Lambda$CDM framework. The constrains from PantheonPlus, DESI BAO, and CMB alone are shown in blue, pink, and orange, respectively. The combined results with FRBs are show in purple and black. All contours show $68\%$ and $95\%$ credible intervals.}
	\label{fig:w0}
\end{figure}

The constraints for the flat $w$CDM model are shown in \autoref{fig:w0}. To verify the reliability of our methodology, we also provide results from individual datasets, which are consistent with those reported by \citet{2024arXiv240403002D}. Notably, the degeneracy between $w$ and $\Omega_m$ can be released when including the FRB dataset, which aligns with predictions of \citet{2014PhRvD..89j7303Z}. The combined analysis yields the following results:
\begin{equation}
\left. \begin{array}{l}
 { {\Omega_m }  = 0.346^{+0.009}_{-0.009}, } \\
 {w   = -1.054^{+0.036}_{-0.037}, }
 \end{array} \right\} \begin{aligned} 
 & {\rm FRBs + PantheonPlus} \\
 & + {\rm DESI \, BAO,} \end{aligned}
\end{equation}
and the tightest constraints are obtained from the combination with CMB dataset:
\begin{equation}
\left. \begin{array}{l}
 { {\Omega_m }  = 0.317^{+0.004}_{-0.004}, } \\
 {w   = -0.991^{+0.021}_{-0.021}, }
 \end{array} \right\} \begin{aligned} 
 & {\rm FRBs + PantheonPlus} \\
 & + {\rm DESI \, BAO + CMB.} \end{aligned}
\end{equation}
Both the estimations of $w$ show consistency with the $\rm \Lambda$CDM framework ($w = -1$). The mild discrepancy between the black and purple contours is due to the dominant influence of CMB dataset. Since the constraints of $\Omega_m$ by CMB tend to be smaller than those from DESI BAO, the combined results align with the CMB posterior contours once CMB dataset is included.

\begin{figure}[ht!]
	\centering
	\includegraphics[width=\linewidth]{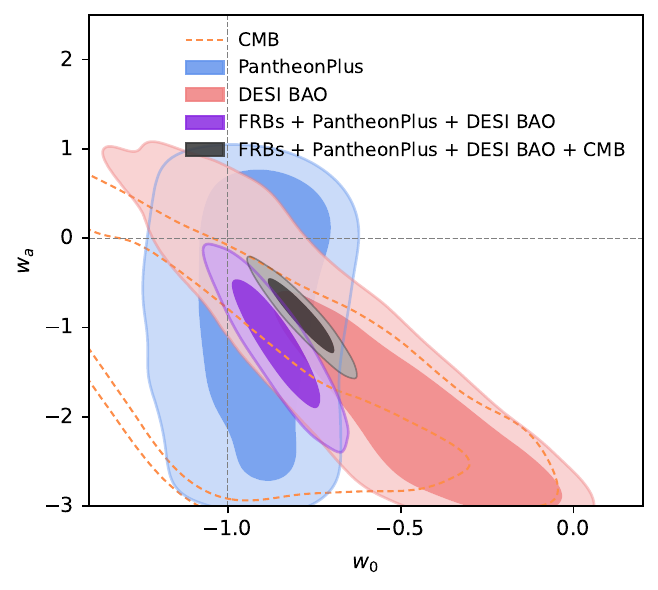}
	\caption{\small The marginalized posterior constraints in the $w_0$ - $w_a$ plane for the flat $w_0 w_a$CDM model. The dashed grey line represents the $\rm \Lambda$CDM framework. The constrains from PantheonPlus, DESI BAO, and CMB alone are shown in blue, pink, and orange, respectively. The combined results with FRBs are show in purple and black. All contours show $68\%$ and $95\%$ credible intervals.}
	\label{fig:w0wa}
\end{figure}

\autoref{fig:w0wa} shows the constraints of flat $w_0 w_a$CDM model. The combined analysis yields:
\begin{equation}
\left. \begin{array}{l}
 { {w_0 }  = -0.855^{+0.084}_{-0.084}, } \\
 {w_a   = -1.174^{+0.462}_{-0.491}, }
 \end{array} \right\} \begin{aligned} 
 & {\rm FRBs + PantheonPlus} \\
 & + {\rm DESI \, BAO.} \end{aligned}
\end{equation}
When incorporating CMB dataset, the constraints are:
\begin{equation}
\left. \begin{array}{l}
 { {w_0 }  = -0.784^{+0.064}_{-0.064}, } \\
 {w_a   = -0.872^{+0.269}_{-0.278}, }
 \end{array} \right\} \begin{aligned} 
 & {\rm FRBs + PantheonPlus} \\
 & + {\rm DESI \, BAO + CMB.} \end{aligned}
\end{equation}
Both results exhibit mild deviations from the standard $\rm \Lambda$CDM at the $\gtrsim 2 \rm \sigma$ credible level, indicating a preference for a present-day quintessence-like EoS ($w_0 >-1$) that crossed the phantom barrier ($w_a <0$). Using the combined dataset of FRBs, PantheonPlus, DESI BAO and CMB, the logarithmic Bayes evidences for the flat $w$CDM and $w_0 w_a$CDM models are $-1899.38$ and $-1895.17$, respectively. Therefore, the Bayes factor is
\begin{equation}
\boldsymbol{\mathcal{B}} = \frac{ P(\boldsymbol{y}|\mathcal{M}_{w_0 w_a {\rm CDM}}) }{ P(\boldsymbol{y}|\mathcal{M}_{w{\rm CDM}}) } = 67.13,
\end{equation}
where $\boldsymbol{y}$ is the dataset, $P$ is the relative evidence of specific model $\mathcal{M}$. As proposed by \citet{Jeffreys:1939xee}, it is a very strong evidence in favor of $w_0 w_a$CDM model, supporting a significant preference for DDE.

\begin{figure*}[ht!]
	\centering
	\includegraphics[width=\linewidth]{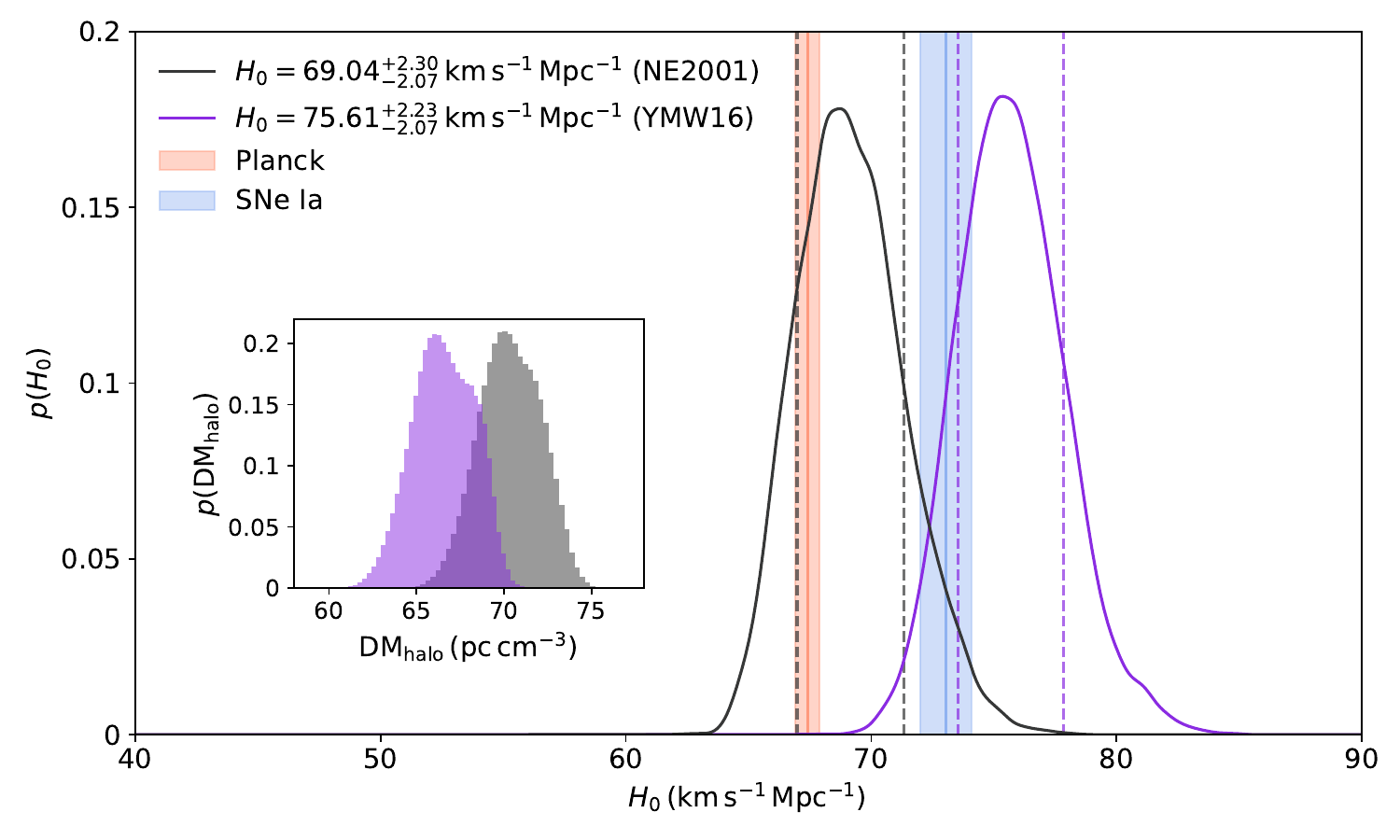}
	\caption{\small The posterior distributions of $H_0$ from localized FRBs only. The black and purple lines represent the cases that assuming NE2001 and YMW16 as the Milky Way electron density models. The orange and blue regions represent the $H_0$ estimations from SNe Ia and CMB, respectively. The inset presents the corresponding distributions of $\rm DM_{halo}$. It shows a collected sample of $\rm DM_{halo}$ for all FRBs. The black and purple regions represent the cases that using NE2001 and YMW16 models, respectively. The slight discrepancy between them is caused by the variation of cosmological parameters.}
	\label{fig:H0}
\end{figure*}

\begin{figure*}[ht!]
	\centering
	\includegraphics[width=\linewidth]{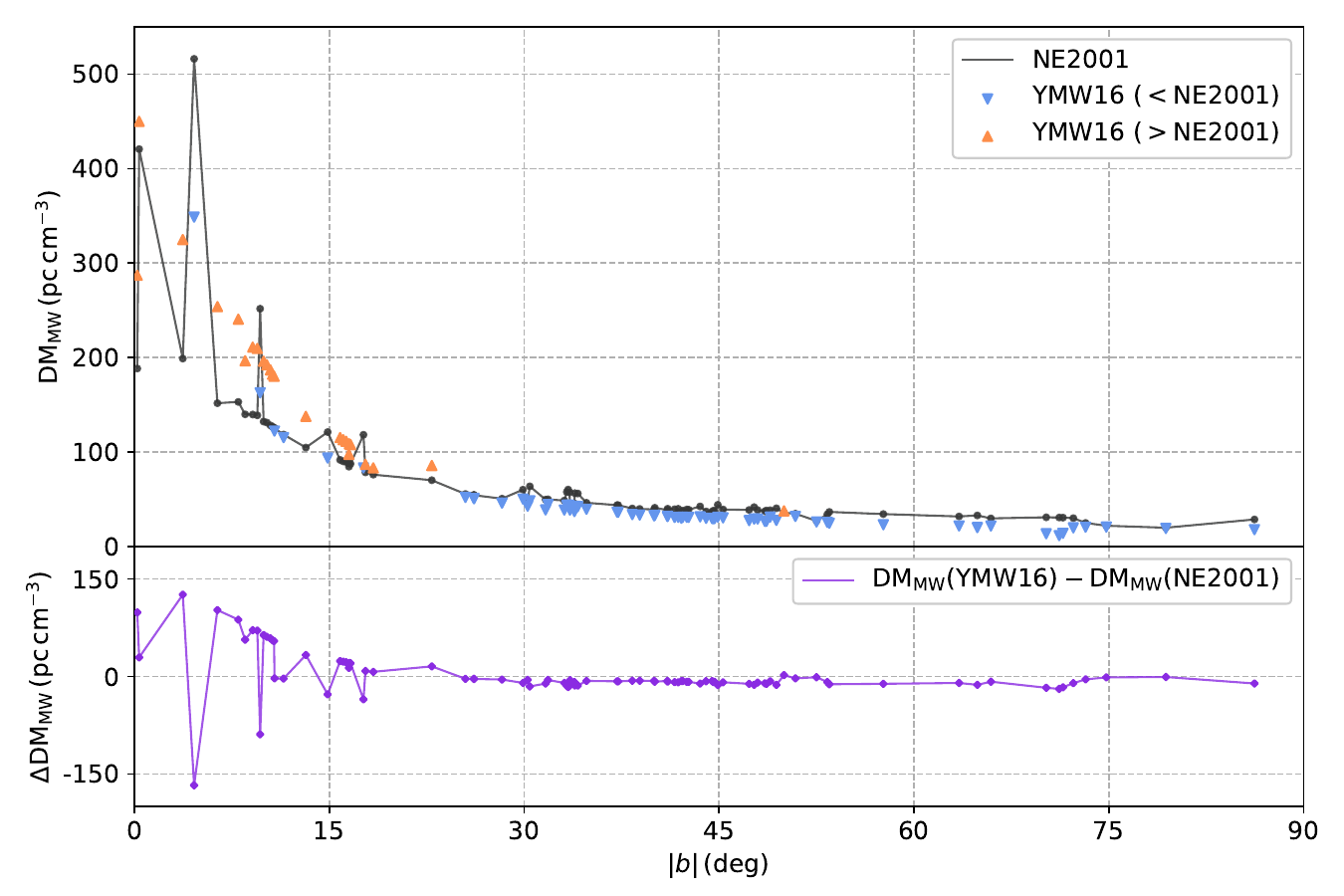}
	\caption{\small Upper panel: $\rm DM_{MW}$ for FRBs using NE2001 and YMW16 models at different Galactic latitude. The black scatters represent the $\rm DM_{MW}$ from NE2001. The orange and blue triangle scatters represent the $\rm DM_{MW}$ from YMW16 with a larger or smaller value compared with NE2001 prospect, respectively. Lower panel: the difference between the $\rm DM_{MW}$ predicted by NE2001 and YMW16.}
	\label{fig:DM}
\end{figure*}

Furthermore, we constrain $H_0$ using updated localized FRBs. The prior of $\Omega_m$ is assumed to follow \citet{2020A&A...641A...6P}, with a Gaussian distribution $\mu=0.315$, $\sigma=0.007$. The priors of $H_0$ and $\Omega_b h^2 $ are consistent with those listed in \autoref{Tab:priors}. Contribution of $\rm DM_{MW}$ to $\rm DM_{IGM}$ estimated by NE2001 and YMW16 models yields slightly distinct results. After reselection with additional criteria $\rm DM_{obs} - DM_{MW} > 80~{\rm pc~cm^{-3}}$, the final number of FRBs are 86 and 88 for NE2001 and YMW16 models, respectively. The constraints of $H_0$ are shown in \autoref{fig:H0}. As shown in \autoref{fig:DM}, most FRBs exhibit accordant $\rm DM_{MW}$ predictions between NE2001 and YMW16, except several outliers localized at low Galactic latitude. Using YMW16 model, we find that two FRBs (FRB201123A and FRB210405I) show significantly lower $\rm DM_{MW}$ values, while approximately 12 FRBs have obvious higher $\rm DM_{MW}$ values compared with NE2001 model.
For localized FRB with confirmed $\rm DM_{obs}$, a smaller $\rm DM_{MW}$ will increase the value of $\rm DM_{IGM} + DM_{halo}$. Since $\rm DM_{IGM}$ is determined by both $\rm DM_{MW}$ and $\rm DM_{halo}$, a higher $H_0$ can be derived as described in \autoref{eq:3}. To illustrate the impact of $\rm DM_{MW}$ model further, we estimate the value of $H_0$ under different scenarios. Defining the difference between two model is,
\begin{equation}
\Delta {\rm DM_{MW}} = \rm DM_{MW, \,NE2001} - DM_{MW, \,YMW16},
\end{equation}
three cases are explored, i.e., case~1: excluding two FRBs with relatively low $\rm DM_{MW}$ ($\Delta {\rm DM_{MW}}> 50 \rm \, pc \, cm^{-3} $) from YMW16 model; case~2: excluding 14 FRBs with apparently distinct $\rm DM_{MW}$ (the absolute value of $\Delta {\rm DM_{MW}}> 50 \rm \, pc \, cm^{-3} $) compared with two MW ISM models
; case~3: restricting the sample to those bursts whose $\rm DM_{MW}$ values are less than $100 \rm \, pc \, cm^{-3}$, suggested by \citet{2020Natur.581..391M}. 

For Case~1, the estimated $H_0$ values are $69.01^{+2.31}_{-2.12}$ and $75.49^{+2.36}_{-2.07} \, \rm km \,s^{-1} \, Mpc^{-1}$ using NE2001 and YMW16 models, respectively. Compared with the estimation without additional selection criteria, the $H_0$ value decreases slightly when using YMW16 model due to the higher $\rm DM_{MW}$ value, which reduces the contribution of $\rm DM_{ISM}$. In Case~2, the remaining FRBs exhibit generally lower $\rm DM_{MW}$, leading to higher $H_0$ values of $70.33^{+2.48}_{-2.28}$ and $76.54^{+2.61}_{-2.45}\rm \, km \, s^{-1} \, Mpc^{-1}$ with NE2001 and YMW16 models, respectively. For Case~3, 20 FRBs are excluded based on a stricter criteria, yielding $H_0 = 63.96^{+2.28}_{-2.13}$ and $65.17^{+2.39}_{-2.04} \rm \, km \, s^{-1} \, Mpc^{-1}$ for the NE2001 and YMW16 models, respectively.

These results indicate that increasing the number of localized FRBs can improve the accuracy of $H_0$ measurement. However, several biases could reduce the accuracy of $H_0$ estimation. In addition to selection criteria and variations in Galactic electron density models, \citet{2024Natur.634.1065B} pointed out that inclination-related bias could significantly underestimate the FRB rate, thus confusing their host environments. If the $\rm DM_{MW}$ values are deemed credible  and an unbiased selection criteria is build, our findings suggest that the Hubble tension could be alleviated with the accumulation of localized FRBs.

To facilitate comparisons with relevant works, \autoref{Tab:compare} presents various estimations of $H_0$ along with the priors of several key parameters. It is apparent that all of the results are contingent upon specific prior assumptions regarding model parameters. Our work adopted a relatively reasonable prior assumption that the values of $\Omega_b h^2$ and $\Omega_{m}$ are acquired from distinct method, and the use of any fixed constant is avoided. Additionally, the $\rm DM_{halo}$ model used in this work is linked to the cosmological framework, thereby reducing the potential bias and enhancing the reliability of our final results.

\begin{deluxetable*}{cccccc}[ht!]
\centering
\tablecaption{The comparisons of $H_0$ estimations by FRBs \label{Tab:compare}}
\tablehead{\colhead{FRBs number}& \colhead{$H_0$} &\colhead{Prior of $\Omega_m$} &\colhead{Prior of $\Omega_b h^2$} &\colhead{$\rm DM_{halo}$ model} &\colhead{Ref.} \\
 &$(\rm km \, s^{-1} \, Mpc^{-1})$ &  & &$({\rm pc \, cm^{-3}})$ & }
\startdata
87 &$69.04^{+2.30}_{-2.07}$ &Gaussian(0.317, 0.007) &Gaussian(0.02233, 0.00036) &Physical model &This work\\
69 &$70.41^{+2.28}_{-2.34}$ &Gaussian(0.311, 0.0056) &Gaussian(0.02242, 0.00014) &65  &\citet{2024arXiv241003994G} \\
64 &$73.41^{+2.28}_{-2.71}$ &0.30966 &$0.04897h^2$ &Uniform(50, 80) &\citet{2024arXiv241001974K}\textsuperscript{c}\\
23 &$69.4^{+4.7}_{-4.7}$ &-\textsuperscript{b} &Gaussian(0.02235, 0.00049) &50 &\citet{2023MNRAS.526.1773F}\\
24 &$95.8^{+7.8}_{-9.2}$ &Uniform(0, 1) &$0.0493h^2$ &Uniform(5, 80) &\citet{2023ApJ...955..101W}\\
18 &$68.81^{+4.99}_{-4.33}$ &Gaussian(0.317, 0.007) &Gaussian(0.02235, 0.00049) &Gaussian(65, 15) &\citet{2022MNRAS.515L...1W}\\
18 &$65.5^{+6.4}_{-5.4}$ &- &$0.0487^{+0.0005}_{-0.0004}h^2$ &$50$ &\citet{2024MNRAS.527.7861G}\\
18\textsuperscript{a} &$71^{+3}_{-3}$ &- &$0.048^{+0.001}_{-0.001}h^2$ &65 &\citet{2023ApJ...946L..49L}\\
16 &$73^{+12}_{-8}$ &0.30966 &0.02242 &50 &\citet{2022MNRAS.516.4862J}\\
9 &$62.3^{+9.1}_{-9.1}$ &$0.143h^2$ &0.2237 &-  &\citet{2022MNRAS.511..662H}\\
\enddata
\tablecomments{\textsuperscript{a} The results from \citet{2023ApJ...946L..49L} considered extra 19 Hubble parameter measurements. Except this work, other works used NE2001 as the Milky way electron density distribution model. \\
\qquad \quad \; \textsuperscript{b} ``-" represents the corresponding parameter was not used in their work.\\
\qquad \quad \; \textsuperscript{c} \citet{2024arXiv241001974K} have discussed various scenarios, so we show the most similar one compared with our work.}
\end{deluxetable*}

\section{Conclusions and Summary}\label{sec:4}
In this work, we combined localized FRBs, DESI BAO, PlantheonPlus, and CMB datasets to constrain the DE EoS using the CPL parameterization. We find no significant deviation from $w=-1$ and the dynamical dark energy component characterized by $w_0>-1$ and $w_a<0$ remains robust. For the flat $w$CDM model, the combined constraints yield $w=-0.991^{+0.021}_{-0.021}$ and $w=-1.054^{+0.036}_{-0.037}$ with and without CMB dataset included, respectively. The Bayes factor comparing $w_0 w_a$CDM and $w$CDM model is $\sim 67.13$, strongly favoring the DDE model with an evolving equation of state parameter $w(z)$. Without CMB dataset, we obtain $w_0 = -0.855 ^{+0.084}_{-0.084}$ and $w_a = -1.174^{+0.462}_{-0.491}$, presenting a $\sim 2.5 \sigma$ tension with $\Lambda$CDM. While including the CMB dataset improves the constraints to $w_0 = -0.784^{+0.064}_{-0.064}$ and $w_a = -0.872^{+0.269}_{-0.278}$, which is consistent with these recent studies \citep{2024arXiv240917074G, 2024arXiv240500502P, 2024arXiv240603395W} using similar datasets. Considering the contribution of FRBs, our results are slightly more restrictive than those of \citet{2024arXiv240403002D}, showing a $\sim 3.1\sigma$ discrepancy with $\Lambda$CDM, compared to their reported $2.5\sigma$ tension using DESI BAO, PantheonPlus, and CMB datasets.

We also investigate the potential of resolving the Hubble tension using localized FRBs alone. Although we derive the tightest $H_0$ constraints, with $69.04^{+2.30}_{-2.07}$ and $75.61^{+2.23}_{-2.07} \, \rm km \, s^{-1} \, Mpc^{-1}$ based on the NE2001 and YMW16 models, these results shows different tendencies, supporting either SNe Ia or CMB measurements. Therefore, a realistic and unified Milky Way electron density model is essential for accurate $H_0$ inference. Without it, the values of $\rm DM_{IGM}$ are always overestimated or underestimated.
For FRBs located in the local universe or behind the Galactic disk, the $\rm DM_{MW}$ contribution dominates, potentially biasing the $\rm DM_{IGM}$ estimation. A possible solution is using high-redshift FRBs, where $\rm DM_{MW}$ contribution are minimal, allowing for a more accurate $\rm DM_{IGM}$ measurement. Furthermore, high-redshift FRBs can clarify whether $f_{\rm IGM}$ evolves with increasing redshift. Although \citet{2020MNRAS.496L..28L} and  \citet{2023MNRAS.520.6237L} found no evidence of redshift dependence for $f_{\rm IGM}$, its value at high redshift require further investigation due to its impact on $H_0$ estimation \citep{2022MNRAS.515L...1W}.

Although our results indicate that FRB observations cannot dominate DE EoS constraints when combined with other datasets, thousands of localized FRBs will be detected in the coming years. \citet{2020ApJ...903...83Z} predicted that 10000 FRBs events with confirmed redshifts could perform better than BAO observations in breaking parameter degeneracies inherent in CMB data. It could constrain DE parameters $w$ to within $5\%$. \citet{2018ApJ...856...65W} constructed mock catalogs with 1000 localized FRBs, showing that although the dark energy equation of state was still poorly constrained, $\Omega_b h^2$ could be determined with a higher precision. Future Square Kilometre Array observations could detect $10^6$ localized FRBs, achieving a $1.6\%$ precision on $w$ independently \citep{2023SCPMA..6620412Z}, surpassing current constraints from CMB, BAO and SNe~Ia.

Besides localized FRBs with reliable host galaxy redshifts, many observed FRBs remain unlocalized. Similar to dark siren method in gravitational wave cosmology, \citet{2022arXiv221213433Z} extracted redshift information from galaxy catalogs to explore cosmological parameters. Despite large uncertainties in current $H_0$ estimations, future detection of unlocalized FRBs will rapidly increase, reducing uncertainties by a factor of $\sqrt{N}$.

In summary, recent BAO observations from DESI reveal a dynamical property of the DE EoS, indicating a feasible orientation to probe cosmological models beyond $\Lambda$CDM. As a promising probe, FRBs can break the degeneracies among cosmological parameters. Future data from BAO measurements from DESI \citep{2019AJ....157..168D}, next-generation CMB experiments \citep{2016arXiv161002743A, 2019JCAP...02..056A} and FRBs from various radio telescope array under construction \citep{2024AstTI...1...84J, 2009IEEEP..97.1482D} will definitely contribute valuable insights into cosmology.

\begin{acknowledgements}
\section{Acknowledgements}
We thank the anonymous referee for helpful comments and suggestions. We thank Yin-Jie Li for his help in developing likelihood functions. This work is supported in part by NSFC under grants of No. 11921003, No. 12233011, No. 12303056 and No. 123B2045, and the Strategic Priority Research Program of the Chinese Academy of Sciences (grant No. XDB0550400).

Software: {\tt Nessai} (\citet{2024ascl.soft05002W}), version 0.12.0, \url{https://nessai.readthedocs.io/en/latest/}. {\tt CLASS} (\citet{2011JCAP...07..034B}), version 3.2.0.1, \url{https://github.com/lesgourg/class_public}. {\tt getdist}, version 1.4.3, \url{https://getdist.readthedocs.io/en/latest/}. {\tt FRB}, verdion 0.1.0, \url{https://github.com/FRBs/FRB}.
\end{acknowledgements}

\clearpage
\appendix
\autoref{Tab:FRBs} presents 92 localized FRBs and their physical properties, including redshift, Right Ascension
(R.A.), Declination (Decl.), $\rm DM$ observations, $\rm DM_{MW}$ predicted by NE2001 and YMW16 models, and the types of their host galaxies.
\startlongtable
\begin{deluxetable}{ccccccccc}
\centering
\tablecaption{92 localized FRBs and their properties\label{Tab:FRBs}}
\tablehead{\colhead{Names} & \colhead{Redshift} & \colhead{$\rm DM_{obs}$} & \colhead{$\rm R.A.$} & \colhead{Decl.} & \colhead{${\rm DM_{MW}}$ (NE2001)} & \colhead{${\rm DM_{MW}}$ (YMW16)} & \colhead{Type} & \colhead{Ref.} \\
& &$(\rm pc \, cm^{-3})$ &(deg, J2000) &(deg, J2000) &$(\rm pc \, cm^{-3})$ &$(\rm pc \, cm^{-3})$ & &}
\startdata
FRB121102A&0.19273&557.0&82.9946&33.1479&188.4&287.1&1 &1,2\\
FRB171020A&0.008672&114.1&333.75&-19.6667&36.7&24.7&3 &4\\
FRB180301A&0.3304&552.0&93.2268&4.6711&151.7&254.0&1 &3\\
FRB180814A&0.068&190.9&65.6833&73.6644&87.6&107.9&2 &16\\
FRB180916B&0.0337&349.349&29.5031&65.7168&199.0&324.9&2 &5\\
FRB180924B&0.3212&361.42&326.1053&-40.9&40.5&27.6&3 &6,8,10,11,25\\
FRB181030A&0.00385&103.396&158.5838&73.7514&41.1&33.0&1 &17\\
FRB181112A&0.4755&589.27&327.3485&-52.9709&41.7&29.0&3 &9,10,11,25\\
FRB181220A&0.02746&208.66&348.6982&48.3421&118.5&115.3&3 &18\\
FRB181223C&0.03024&111.61&180.9207&27.5476&19.9&19.1&3 &18\\
FRB190102C&0.2912&364.5&322.4157&-79.4757&57.4&43.3&3 &10,11,25\\
FRB190110C&0.12244&221.6&249.3185&41.4434&37.1&29.9&2 &19\\
FRB190303A&0.064&223.2&207.9958&48.1211&29.8&21.8&2 &16\\
FRB190418A&0.07132&182.78&65.8123&16.0738&70.2&85.8&3 &18\\
FRB190425A&0.03122&127.78&255.6625&21.5767&48.7&38.7&3 &18\\
FRB190520B&0.2418&1204.7&240.5178&-11.2881&60.2&50.2&1 &6,12\\
FRB190523A&0.66&760.8&207.065&72.4697&37.2&29.9&3 &20\\
FRB190608B&0.11778&338.7&334.0199&-7.8983&37.3&26.6&3 &6,10,11,25\\
FRB190611B&0.3778&321.4&320.7456&-79.3976&57.8&43.7&3 &7,10,25\\
FRB190614D&0.6&959.2&65.0755&73.7067&87.8&108.7&3 &21\\
FRB190711A&0.522&593.1&329.4193&-80.358&56.5&42.6&1 &7,10,25\\
FRB190714A&0.2365&504.13&183.9797&-13.021&38.5&31.2&3 &7,25\\
FRB191001A&0.234&506.92&323.3513&-54.7478&44.2&31.1&3 &7,25\\
FRB191106C&0.10775&332.2&199.5801&42.9997&25.0&20.5&2 &19\\
FRB191228A&0.2432&297.5&344.4304&-28.5941&32.9&20.1&3 &3,25\\
FRB200223B&0.06024&201.8&8.2695&28.8313&45.6&37.0&2 &19\\
FRB200430A&0.1608&380.1&229.7064&12.3763&27.2&26.1&3 &7,25\\
FRB200906A&0.3688&577.8&53.4962&-14.0832&35.8&37.9&3 &3,25\\
FRB201123A&0.0507&433.55&263.67&-50.76&251.7&162.7&1 &22\\
FRB201124A&0.098&413.52&77.0146&26.0607&139.9&196.6&2 &13\\
FRB210117A&0.2145&729.1&339.9792&-16.1515&34.4&23.1&3 &6,25\\
FRB210320C&0.2797&384.8&204.4608&-16.1227&39.3&30.4&3 &6,25\\
FRB210405I&0.066&565.17&255.3396&-49.5452&516.1&348.7&3 &26\\
FRB210410D&0.1415&571.2&326.0863&-79.3182&56.2&42.2&3 &6,14\\
FRB210603A&0.1772&500.147&10.2741&21.2263&39.5&30.8&3 &23\\
FRB210807D&0.1293&251.9&299.2214&-0.7624&121.2&93.7&3 &6,25\\
FRB211127I&0.0469&234.83&199.8082&-18.8378&42.5&31.5&3 &6,15,25\\
FRB211203C&0.3439&636.2&204.5625&-31.3801&63.7&48.4&3 &6,25\\
FRB211212A&0.0707&206.0&157.3509&1.3609&38.8&27.5&3 &6,25\\
FRB220105A&0.2785&583.0&208.8039&22.4665&22.0&20.6&3 &6,25\\
FRB220204A&0.4012&612.584&274.2263&69.7225&50.7&46.0&3 &24,28,29\\
FRB220207C&0.04304&262.38&310.1995&72.8823&76.1&83.3&3 &24,29\\
FRB220307B&0.248123&499.27&350.8745&72.1924&128.2&186.9&3 &24,29\\
FRB220310F&0.477958&462.24&134.7204&73.4908&46.3&39.5&3 &24,29\\
FRB220319D&0.011228&110.98&32.1779&71.0353&139.7&211.0&3 &24\\
FRB220418A&0.622&623.25&219.1056&70.0959&36.7&29.5&3 &24,29\\
FRB220501C&0.381&449.5&352.3792&-32.4907&30.6&14.0&3 &25,28\\
FRB220506D&0.30039&396.97&318.0448&72.8273&84.6&97.7&3 &24,28,29\\
FRB220509G&0.0894&269.53&282.67&70.2438&55.6&52.1&3 &18,24,29\\
FRB220529A&0.1839&246.0&19.1042&20.6325&40.0&30.9&1 &27\\
FRB220610A&1.016&1458.15&351.0732&-33.5137&31.0&13.6&3 &25\\
FRB220717A&0.36295&637.34&293.3042&-19.2877&118.3&83.2&3 &32\\
FRB220725A&0.1926&290.4&353.3152&-35.9902&30.7&11.6&3 &25\\
FRB220726A&0.3619&686.232&73.94567&69.9291&89.5&111.4&3 &24,28,29\\
FRB220825A&0.241397&651.24&311.9815&72.585&78.5&86.9&3 &24,29\\
FRB220831A&0.262&1146.25&338.6955&70.5384&126.8&182.3&3 &29\\
FRB220912A&0.0771&219.46&347.2704&48.7071&125.2&122.2&2 &31\\
FRB220914A&0.1139&631.28&282.0568&73.3369&54.7&51.1&3 &24,29\\
FRB220918A&0.491&656.8&17.5921&70.8113&153.1&240.6&3 &25\\
FRB220920A&0.158239&314.99&240.2571&70.9188&39.9&33.4&3 &24,29\\
FRB221012A&0.284669&441.08&280.7987&70.5242&54.3&50.5&3 &24,29\\
FRB221029A&0.975&1391.75&141.9634&72.4523&43.8&36.4&3 &24,28,29\\
FRB221101B&0.2395&491.554&342.2162&70.6812&131.2&192.4&3 &24,28,29\\
FRB221106A&0.2044&343.8&56.7048&-25.5698&34.8&31.8&3 &24,25\\
FRB221113A&0.2505&411.027&71.411&70.3074&91.7&115.4&3 &24,28,29\\
FRB221116A&0.2764&643.448&21.2102&72.6539&132.3&196.2&3 &28,29\\
FRB221219A&0.553&706.708&257.6298&71.6268&44.4&38.6&3 &24,28,29\\
FRB230124A&0.0939&590.574&231.9163&70.9681&38.6&31.8&3 &24,28,29\\
FRB230307A&0.2706&608.854&177.7813&71.6956&37.6&29.5&3 &24,28,29\\
FRB230501A&0.3015&532.471&340.0272&70.9222&125.7&180.2&3 &24,29\\
FRB230521B&1.354&1342.9&351.036&71.138&138.8&209.7&3 &25,29\\
FRB230526A&0.157&361.4&22.2326&-52.7173&31.9&21.9&3 &25\\
FRB230626A&0.327&452.723&235.6296&71.1335&39.3&32.5&3 &24,28,29\\
FRB230628A&0.127&344.952&166.7867&72.2818&39.0&30.8&3 &24,28,29\\
FRB230708A&0.105&411.51&303.1155&-55.3563&60.3&44.0&3 &25\\
FRB230712A&0.4525&587.567&167.3585&72.5578&39.2&30.9&3 &24,28,29\\
FRB230718A&0.0357&477.0&128.1619&-40.4519&420.6&450.0&3 &25\\
FRB230814A&0.553&696.4&335.9748&73.0259&104.8&137.8&3 &29\\
FRB230902A&0.3619&440.1&52.1398&-47.3335&34.1&25.5&3 &25\\
FRB231120A&0.0368&437.737&143.984&73.2847&43.8&36.2&3 &24,28,29\\
FRB231123B&0.2621&396.857&242.5382&70.7851&40.3&33.8&3 &24,28,29\\
FRB231220A&0.3355&491.2&123.9087&73.6599&49.9&44.5&3 &29\\
FRB231226A&0.1569&329.9&155.3638&6.1103&38.1&26.7&3 &25\\
FRB240114A&0.13&527.65&321.9161&4.3292&49.7&38.8&1 &30\\
FRB240119A&0.376&483.1&224.4672&71.6118&38.0&31.0&3 &29\\
FRB240123A&0.968&1462.0&68.2625&71.9453&90.2&113.0&3 &29\\
FRB240201A&0.042729&374.5&149.9056&14.088&38.6&29.1&3 &25\\
FRB240210A&0.023686&283.73&8.7796&-28.2708&28.7&17.9&3 &25\\
FRB240213A&0.1185&357.4&166.1683&74.0754&40.0&32.1&3 &29\\
FRB240215A&0.21&549.5&268.4413&70.2324&47.9&42.8&3 &29\\
FRB240229A&0.287&491.15&169.9835&70.6762&38.0&29.5&3 &29\\
FRB240310A&0.127&601.8&17.6219&-44.4394&30.1&19.8&3 &25\\
\enddata

\tablecomments{References: 1, \citet{2017Natur.541...58C}, 2, \citet{2017ApJ...834L...7T}, 3, \citet{2022AJ....163...69B}, 4, \citet{2018ApJ...867L..10M}, 5, \citet{2020Natur.577..190M}, 6, \citet{2023ApJ...954...80G}, 7, \citet{2020ApJ...903..152H}, 8, 
\citet{2019Sci...365..565B}, 9, \citet{2019Sci...366..231P}, 10, \citet{2020Natur.581..391M}, 11, \citet{2020ApJ...895L..37B}, 12, \citet{2022Natur.606..873N}, 13, \citet{2021ApJ...919L..23F}, 14, \citet{2023MNRAS.524.2064C}, 15, \citet{2023ApJ...949...25G}, 16, \citet{2023ApJ...950..134M}, 17, \citet{2021ApJ...919L..24B}, 18, 
\citet{2024ApJ...971L..51B}, 19, \citet{2024ApJ...961...99I}, 20, \citet{2019Natur.572..352R}, 21, \citet{2020ApJ...899..161L}, 22, \citet{2022MNRAS.514.1961R}, 23, \citet{2023arXiv230709502C}, 24, \citet{2024ApJ...964..131S}, 25, \citet{2024arXiv240802083S}, 26, \citet{2024MNRAS.527.3659D}, 27, \citet{2024arXiv241003994G}, 28, 
\citet{2024Natur.635...61S}, 29, \citet{2024arXiv240916952C}, 30, \citet{2024MNRAS.533.3174T}, 31, \citet{2023ApJ...949L...3R}, 32, \citet{2024MNRAS.532.3881R} \\
This table is available in machine-readable format from the online journal.}
\end{deluxetable}

\clearpage
\bibliography{ref}
\bibliographystyle{aasjournal}
\end{document}